\documentclass[10pt,aps,pra,twocolumn,amsmath,amssymb,floatfix,superscriptaddress]{revtex4-2}
\usepackage{graphicx}
\usepackage[normalem]{ulem}
\usepackage[title]{appendix}
\usepackage{hyperref}
\usepackage{comment}

\newcommand{\ket}[1]{\lvert #1 \rangle}
\newcommand{\bra}[1]{\langle #1 \rvert}
\newcommand{\ketbra}[2]{\ket{#1}\bra{#2}}

\usepackage[usenames,dvipsnames]{color}

\newcommand{\affA}{Van der Waals-Zeeman Institute, Institute of Physics, University of Amsterdam, 1098 XH Amsterdam, Netherlands}
\newcommand{\affB}{QuSoft, Science Park 123, 1098 XG Amsterdam, the Netherlands}

\begin{document}

\title[Sample title]{Alignment and Optimisation of Optical Tweezers on Trapped Ions}
\author{M. Mazzanti}\affiliation{\affA}
\author{C. Robalo Pereira}\affiliation{\affA} 
\author{N. A. Diepeveen}\affiliation{\affA}
\author{B. Gerritsen}\affiliation{\affA}\affiliation{\affB}
\author{Z. Wu}\affiliation{\affA}
\author{Z. E. D. Ackerman}\affiliation{\affA}
\author{L. P. H. Gallagher}\affiliation{\affA}
\author{A. Safavi-Naini}\affiliation{\affA}\affiliation{\affB}
\author{R. Gerritsma}\affiliation{\affA}\affiliation{\affB}
\author{R. X. Schüssler}\affiliation{\affA}

\begin{abstract}

This paper presents a routine to align an optical tweezer on a single trapped ion 
and use the ion as a probe to characterize the tweezer.
We find a smallest tweezer waist of $2.3(2)\,\mu$m, which is in agreement with  the theoretical minimal attainable waist of $2.5(2)\,\mu$m in our setup. 
We characterize the spatial dependence of the tweezer Rabi frequency which is suppressed by a factor of 19(3) in the immediate surrounding of the ion. We investigate the effects of optical forces and coherent population trapping on the ion. Finally, we show that the challenges posed by these forces can be overcome, and that the number of tweezers can be easily scaled up to reach several ions by using a spatial light modulator.

\end{abstract}

\date{\today}
\maketitle

\section{Introduction}
\label{sec:level1}
The precise control offered by optical tweezers has led to significant advances in atomic and quantum physics. For instance, atoms trapped in optical tweezers have been used to study phenomena such as quantum phase transitions, superfluidity, and quantum magnetism \cite{Grier:2003,Scholl:2021,Bluvstein:2021,Omran:2019,Ashkin:1986, Beugnon:2007}, as well as offering significant potential for quantum computing by providing precise control over atom arrangements and system connectivity \cite{Barredo:2016, Endres:2016, Omran:2019, Graham:2019, Kaufman:2021,Urech:2022,bluvstein:2024}. 

More recently, it has been shown that combining optical tweezers with trapped ions allows for new quantum computing architectures as well as more flexible quantum simulation platforms~\cite{ Teoh:2021, Espinoza:2021, Nath:2015, Olsacher:2020, Bond:2022, Mazzanti:2021}. 
In these platforms, individual addressing of ions, typically separated by a few micrometers in an ion crystal, is achieved using tightly focused laser beams. Although these addressing beams are not used to confine the ions, they allow for enhanced control over interactions by locally manipulating ion confinement and, consequently, the phonon modes that mediate the interactions between the ions~\cite{Espinoza:2021, Bond:2022}.
However, there are several challenges that must first be overcome. First, one needs to supply optical tweezers with sufficient intensity to compete with the electric potentials of the ion trap. Second, it is crucial to reliably align the optical tweezers onto the trapped ions while compensating for aberrations in the tweezer delivery system.
Techniques for optimizing the tweezer potential are more established in neutral atom platforms, ~\cite{Cizmar:2010,Stilgoe:2021,Schulze:2013,Mochi:2015,Booth:2002,Debarre:2009} as opposed to trapped ion setups.
Recent developments on trapped ions include the application of hollow tweezers ~\cite{Schmiegelow:2016,Drechsler:2021,Stopp_ang_mom2022}, as well as programmable holographic beams implemented with a Digital Micromirror Device~\cite{Shih:2021}. 


\begin{figure}[t]
\includegraphics[width=\linewidth]{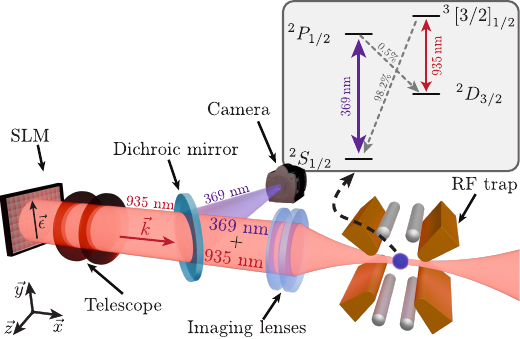}
\caption{Experimental setup with RF trap, SLM, tweezers telescope and imaging lenses. 935\,nm light with polarization $\epsilon$ and wavevector $\vec{k}$ is reflected by a SLM and focused at the ion's position.The fluorescence light emitted by the ion is separated using a dichroic mirror and imaged on a camera. In the inset are shown the relevant $^{174}$Yb$^{+}$ transitions.
A combination of rod  (grey) and endcaps (not shown in figure) electrodes allow to change the ion's position in order to map the beam's intensity profile.}
\label{fig:expsetup}
\end{figure}
In this paper, we demonstrate the alignment and optimization of optical tweezers on $^{174}$Yb$^+$ ions trapped in a radiofrequency (RF) trap. 
From our measurements we model the tweezer to extract the waist and demonstrate the optimization the tweezer waist to the theoretical limit.  We investigate the interaction of the trapped ion with a high power tweezer close to resonance, and observe the effects of optical forces and coherent population trapping. 
We develop a theoretical model that explains the spatial dependence of the Rabi frequency from our data.
We find a maximum Rabi frequency of $\Omega_{\rm tw}=2\pi\times 390(70)$\, MHz and a background Rabi frequency suppression factor of $19(3)$ in the immediate vicinity of the ion. Finally, we program our SLM to deliver two optimized tweezers on two ions in a 5-ion crystal, demonstrating the scalability of the routine. 








\begin{figure*}[t]
    \includegraphics[trim={0.6cm 0 0 0},width=\linewidth]{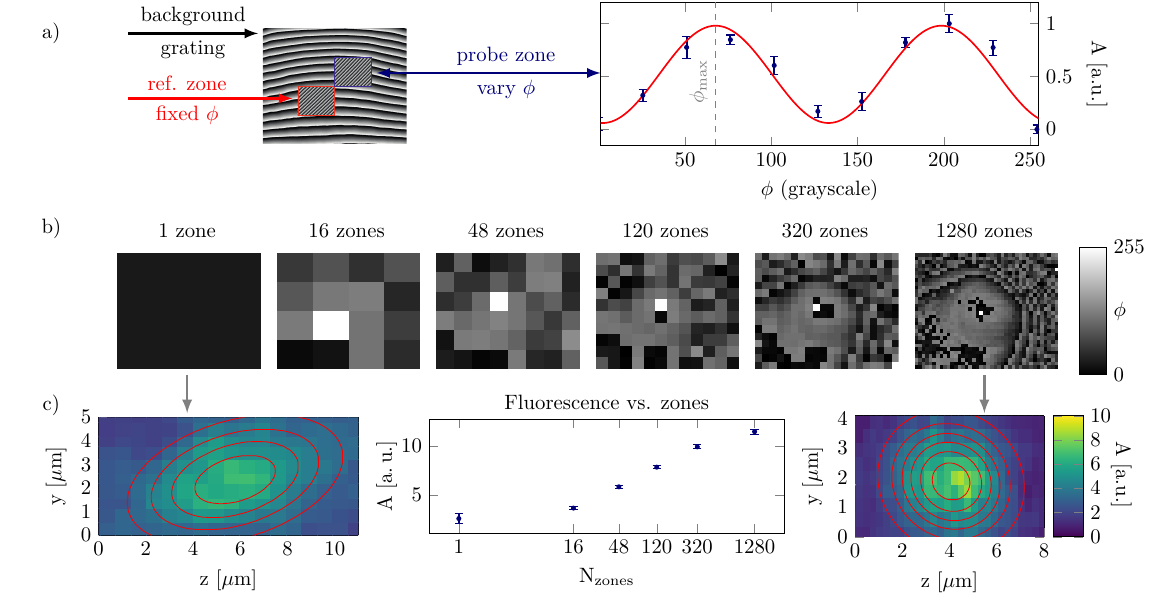}
    \caption{Tweezers optimisation algorithm. a) The SLM screen is divided into zones, with one taken as a reference zone with a fixed phase $\phi$. One other zone is used as a probe. The ion fluorescence is recorded as a function of the phase of the probe zone. The phase at which the fluorescence is maximal is used for the aberration correct for this zone. A background grating steers the rest of the beam away from the ion.
    b) Aberration correction patterns for different number N of zones. 
    c) Result of the aberration correction. The fluorescence of the ion signal increases by a factor of 4.4(0.9) from no aberration correction (1 zone) to 1280 zones (middle). The unoptimised beam (left) also shows an oval shape with waists of  $w_1 = 6.1(4)\,\mu$m and $w_2 = 3.2(2)\,\mu$m, while the fully optimised is nearly circular ($w_1 = 2.9(2)\,\mu$m and $w_2 = 2.3(2)\,\mu$m,  right). Both beam characterisations are shown with the same scale.
    }
    \label{fig:zones}
\end{figure*}

\section{Setup}

As shown in Fig.~\ref{fig:expsetup}, we trap a single $^{174}$Yb$^+$ ion in a Paul trap with a RF frequency of 5.8\,MHz and secular trap frequencies of $\omega_{x,y,z}\simeq 2\pi\times(400, 400, 120)$\,kHz in the two radial and axial directions respectively.
We use the $^2S_{1/2}\rightarrow~^2P_{1/2}$ transition at 369\,nm to Doppler cool the ion to the center of the trap.
The $^2P_{1/2}$ state decays with a branching fraction of $0.5\%$ to the $^2D_{3/2}$ state, which has a lifetime of $54.84$\,ms~\cite{Shao:2023}. We optically pump the population in the $^2D_{3/2}$ state back to the cooling cycle using 935\,nm light that couples to the $^3\left[3/2\right]_{1/2}$ state, whose decay has a branching fraction of $98.2\%$ back to the $^2S_{1/2}$ state~\cite{Olmschenk:2007}. 



We use a 2-inch doublet lens system with a numerical aperture (NA) of 0.3, achieving a magnification of $M=8.54(5)$, to collect the ion fluorescence. 
The 369\,nm light is reflected by a dichroic mirror and directed onto a CCD camera for imaging purposes.
A second repumper laser at 935\,nm is used as the tweezer beam and is overlapped with the fluorescence of the ion at the dichroic mirror. The tweezer beam is magnified from a size that nearly spans the entire surface of the SLM ($16\times12.8\,\mathrm{mm}^2$) to a 2-inch diameter using a two-lens telescope system. 
It is then focused down to the ions using the same doublet lens system used in the fluorescence collection.
Our numerical simulation of the entire optical system indicates that the theoretical minimum achievable tweezer waist is $2.5(2)\,\mu$m at 935\,nm.


\section{Tweezer alignment and optimisation}


\noindent \emph{ Alignment and mapping the tweezer:} The time the ion spends in the $^2D_{3/2}$ state, and consequently its fluorescence, depends on the beam intensity.
We determine the intensity profile of the 935\,nm tweezer by moving the ion through the beam along the $y$ and $z$ directions, adjusting the voltage on the compensation and endcap electrodes, respectively. Although the resulting fluorescence image of the ion is $\sim 5\,\mu$m wide, we achieve higher resolution in measuring the tweezer waist by calibrating the ion's location for each applied voltage. To achieve this, we minimize the ion's micromotion amplitude to $\lesssim 0.1\,\mu$m and use a second 935\,nm beam with a large enough diameter to ensure the ion's fluorescence is always at its maximum. We fit the fluorescence image using a two-dimensional Gaussian.
We note that, at a Doppler limited temperature $T\sim 1$\,mK, the spread of the ion position due to its thermal motion, is $\sqrt{k_\text{B}T/(m\omega_z^2)} \approx 232$\,nm, where $k_\text{B}$ is the Boltzmann constant and $m$ is the ion's mass.
By combining the ion's position at a given endcap voltage with the fluorescence image of the ion in the tweezer, we can probe the tweezer light field with sub-waist resolution. Two such maps are shown in Fig.~\ref{fig:zones}(c).

\noindent\emph{Aberration correction:} We create the desired tweezer pattern on the ion by reflecting the tweezer beam off the SLM. The SLM imprints a phase pattern, as shown in Fig.~\ref{fig:zones}(a), onto the wavefront of the tweezer beam. The beam is focused on the ion making use of an appropriate diffraction grating pattern as described in App.~\ref{appn:C}.
As the beam passes through the elements of the optical system, imperfections are introduced both at the SLM and by the different optical elements. The flatness and defects of the SLM chips are corrected using a flatness correction pattern provided by the manufacturer while an additional phase pattern corrects the aberrations introduced by the optical elements. The latter pattern is obtained using the algorithm described in Refs.~\cite{Cizmar:2010, Stilgoe:2021}. The SLM pixels are divided into a number of zones (see Fig. \ref{fig:zones}(a)) with one zone serving as the reference, and others acting as probes. First, we apply the same diffraction grating pattern to the reference and the selected probe zone and a different diffraction grating pattern to the other zones to reduce background fluorescence. This ensures that the scattered light does not converge at the ion's position. We then vary the global phase $\phi$ of the probe zone and record the fluorescence of the ion to obtain the interference pattern shown in Fig.~\ref{fig:zones} (a) and adjust the laser power at the SLM to prevent the saturation of the ion's transition. At the point of maximal fluorescence, the global phase $\phi_{\text{max}}$ of the probe zone cancels out any unwanted phase shifts due to aberrations in the beam path. We repeat this procedure for each probe zone  and find the aberration correction patterns in Fig.~\ref{fig:zones} (b) where the reference zone is the white square at the center. 

\begin{figure}[t]
    \includegraphics[trim={0.6cm 0 0 0},width=\linewidth]{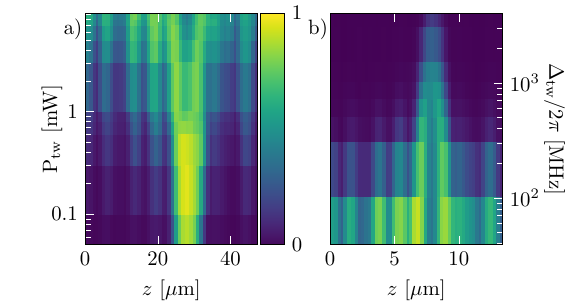}
    \caption{Effect of coherent population trapping on the ion fluorescence as a function of (a) ion position and power $P_{\text{tw}}$ in the beam (taken at $\Delta_{\text{tw}}=2\pi\times 40\textrm{\,MHz}$), and (b) detuning $\Delta_{\text{tw}}/2\pi$ (taken at $P_{\text{tw}}=$ 12.5\,mW). 
    }
    \label{fig:cpt2}
\end{figure}

We show the effect of the aberration correction procedure on the fluorescence of the ion and the tweezer beam shape in Fig.~\ref{fig:zones}(c). In the absence of any correction (1 zone), the beam has an elliptical shape with beam waists of $w_1 = 6.1(4)\,\mu$m and $w_2 = 3.2(2)\,\mu$m for the major and minor axes of the ellipse. 
We perform the procedure described above for an increasing number of zones to achieve a nearly circular beam shape with  waists of $w_1 = 2.9(2)\,\mu$m and $w_2 = 2.3(2)\,\mu$m, while simultaneously increasing the fluorescence and therefore tweezer intensity at the position of the ion. We note that while limited laser power restricts us to 1280 zones, in light of the flattening of the fluorescence curve in Fig.~\ref{fig:zones}(c), any further improvement provided by increasing the number of zones will be small. 

We determine $w_1$ and $w_2$ from the fluorescence measurements using a theoretical model detailed in Appendix~\ref{appn:A}. This model allows us to derive the tweezer waist by incorporating a position-dependent $\Omega_{\rm tw}(\mathbf{r})$. We use optical Bloch equations to calculate the ion's fluorescence in the tweezer, assuming that the ion is a stationary point source with four internal levels. We calculate the scattering rate on the Doppler transition and find, 
\begin{equation}
\resizebox{0.9\columnwidth}{!}{$
    \Gamma_{\rm sc}=\frac{\Gamma_{\text{S,}3/2} \Gamma_{\text{P}} \Omega_{\rm D}^2 \Omega_{\rm tw}^2}{\Gamma_{\text{S,}3/2} \Omega_{\rm tw}^2 \left( \Gamma_{\text{P}}^2 + 2 \Omega_{\rm D}^2 + 4 \Delta_{\rm D}^2 \right)+ \Gamma_{\text{S,P}} \Omega_{\rm D}^2 \left( \Gamma_{3/2}^2 + 2 \Omega_{\rm tw}^2 + 4 \Delta_{\rm tw}^2 \right)},
    $}
\label{eqn:4levelscattering}
\end{equation}
where $\Omega_{\text{D}},~\Delta_{\text{D}}$ are the Rabi frequency and detunings of the Doppler beam, respectively, and $\Delta_{\text{tw}}$ is the detuning of the tweezer. The relevant decay rates $\Gamma$ are given in the appendix.

\begin{figure}[t]
     \centering
     \includegraphics[width=\linewidth]{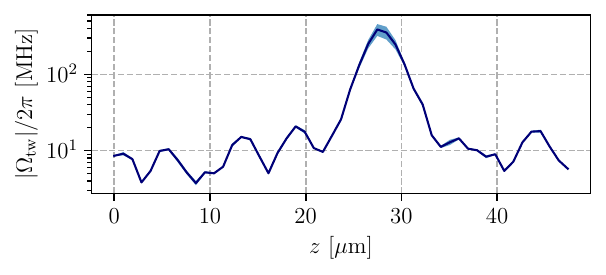}
     \caption{Tweezer Rabi frequency along the axial direction of the Paul trap obtained by numerically fitting the 10-level model to the data shown in Fig.~\ref{fig:cpt2}.}
     \label{fig:rabifit}
\end{figure}

We further characterize the tweezer by studying fluorescence of the ion as a function of the tweezer laser power and its detuning from resonance (see Fig.~\ref{fig:cpt2}) and map the tweezer along the axis of the Paul trap using the procedure described in the previous section. The tweezer is well localised at low power and large detuning. 
As the tweezer power is increased and the detuning is decreased, the background light scattered by the SLM that is outside the central tweezer spot becomes visible. Moreover, we find that while the fluorescence of the ion first increases with larger laser power and smaller detuning, it later decreases. We attribute this effect to coherent population trapping~\cite{Ejtemaee:2010} as the tweezer drives the $J=3/2\to J'=1/2$ transition, in which there are two dark states. We model the data in Fig.~\ref{fig:cpt2} using the optical Bloch equations to calculate the ion's fluorescence in the tweezer. Here, we again assume the ion to be a stationary point source but now with 10 internal levels (see Appendix~\ref{appn:A} for more details) and calculate the scattering rate numerically. Next we use a numerical fit function $f=a\Gamma_{sc}(\Omega_{\text{tw}},\Delta_{\text{tw}})+ b$ to extract $\Omega_{\rm tw}$ at each distance $z$ and as a function of the tweezer detuning $\Delta_{\rm tw}$, by fitting the experimental data in Fig.~\ref{fig:cpt2}(b).
We show the results in Fig.~\ref{fig:rabifit} where we use the same fitted value for the background $b$ for all positions, while allowing the amplitude $a$ to vary. 
We find a peak $\Omega_{\rm tw} = 2\pi \times 390 (70)$ MHz at the center of the tweezer, which is $19(3) \times$ larger than the Rabi frequency for the largest background peak.

Finally, we note that the coherent population trapping may hinder the aberration correction procedure as it may lead to misalignment of the tweezer when the fluorescence of the ion goes down with increasing tweezer intensity. This problem can be avoided by using far detuned and low power tweezer light.



\begin{figure}[t]
    \centering
    \includegraphics[width=\linewidth]{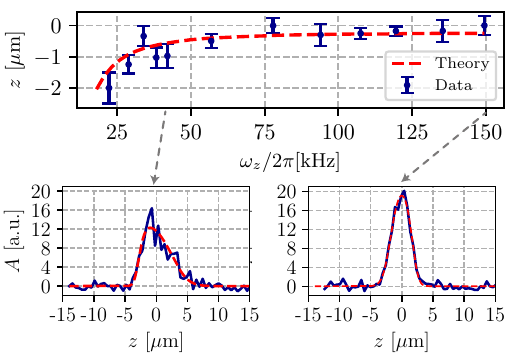}
    \caption{Effect of radiation pressure on the characterization of the tweezers.
    Top: The center of the fit to the tweezers map shifts as a function of the $\omega_z$. The model for the theory curve is given in the appendix.
    Bottom: Fluorescence as a function of ion position for an $\omega_z= 2\pi \times 38$\,kHz (left) and of $\omega_z= 2\pi \times150$\,kHz (right).}
    
    \label{fig:radiationP}
\end{figure}

\section{Effect of optical forces on alignment}
So far we have approximated the ion as a stationary point particle located at the equilibrium position set by the electrode voltages. Here we study the effect of additional optical forces on the ion, for example from the Doppler beam, on the tweezer optimization and mapping procedure described previously. 
To this end, we move the ion through the tweezer to map the beam at different axial confinements
(see Fig. \ref{fig:radiationP}). We find that at $\omega_z$ below $2\pi \times 60$\,kHz the radiation pressure from the Doppler beam pushes the ion out of the center of the tweezer, decreasing the count rate, as well as making the map of the beam appear more asymmetric and less focused. As a result, using our fit procedure with a skew Gaussian function, it appears that the center of the tweezer is shifted by about $1.8(3)\,\mu$m at $\omega_z= 2\pi \times 38$\,kHz. For larger trapping frequencies, the mapped tweezer shape is Gaussian and the extracted value for the center does not change with the trap frequency. We note that all tweezer characterization measurements and aberration correction results presented earlier were done at an $\omega_z= 2\pi \times 120$\,kHz to circumvent this effect.

\section{Two tweezers}

We generate two tweezers by partitioning the SLM using a checkerboard pattern. 
One section contains a grating to target one ion, while the other section targets a different ion. 
The grating settings were found beforehand using a single tweezer.
To have no additional discrete phase jumps the checkerboard rectangles have the same size as the aberration correction zones for the 1280 zones.

\begin{figure}[t]
    \centering
    \includegraphics[trim={1cm 0 0 0},width=0.95\linewidth]{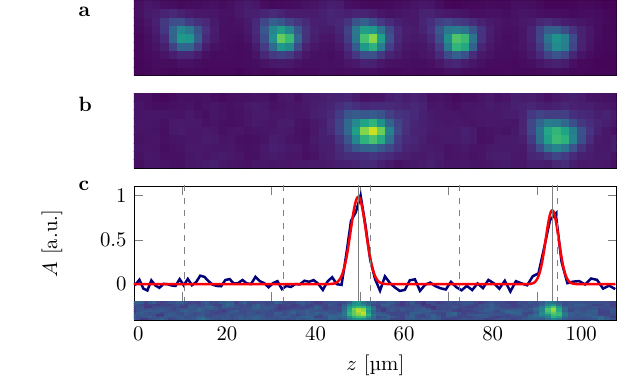}
    \caption{Addressing of individual ions with two tweezers.  a) Picture of a five-ion crystal. b) Same crystal, but only the middle and the outer right ions are addressed by tweezers, making them fluoresce.
    c) Characterization of the tweezer as a plot of fluorescence against ion position and the two tweezers shape.}
    \label{fig:twotweezers}
\end{figure}

In Fig. \ref{fig:twotweezers}a), we show a five-ion crystal. The variations in fluorescence result from the finite waist of the 369\,nm Doppler beam. 
By turning off the second resonant 935\,nm repumper beam and activating the two tweezers with the checkerboard pattern we observe the fluorescence of only two ions, as seen in Fig. \ref{fig:twotweezers}(b), indicating the presence of two tweezers. 

We map the two tweezers by moving one ion across the extent of the 5 ion crystal as shown in Fig.~\ref{fig:twotweezers}(c).
The gray dashed lines indicate the equilibrium position of the five ions. 
The mapping reveals that there is fluorescence only on the two desired ions, with no signal observed on the other ions or in between them. 
We notice the misalignment of the tweezers relative to the ion center positions when we perform the full mapping as the effect is not present in the two-ion picture.  
The misalignment can be fixed by changing the grating of the SLM pattern or changing the inter-ion distance and position with the voltages applied to the trap.
We determined the waists of the left tweezer as $w_1 = 3.4(2)\,\mu$m and $w_2 = 2.5(2)\,\mu$m and those of the right tweezer as $w_1 = 2.8(3)\,\mu$m  and  $w_2 = 2.3(2)\,\mu$m. For more advanced tweezer patterns the zones can be grouped differently or the Gerchberg-Saxton (GS) algorithm can be used~\cite{Gerchberg:1972,DiLeonardo:2007}.







\section{Conclusions and outlook}\label{Sec_conclusions}

We presented a method for aligning an optical tweezer on a single trapped ion. To this end, we developed a protocol to minimize the aberrations introduced by the optical system and the SLM, and to optimize the tweezer light intensity at the location of the ion. in this way we were able to obtain a tweezer waist on the order of the diffraction limit of our system. Moreover, we demonstrated that after aberration correction the SLM can be reprogrammed to create more tweezers. These tweezers may be used to modify the local confinement of the ions and alter their phonon modes and frequencies. This is a crucial step to program trapped ion quantum simulators as proposed in Refs.~\cite{Espinoza:2021,Bond:2022}.

The optical tweezers used in this study did not have enough power to make significant changes to the confinement of the ions. Infrared lasers at wavelengths $>$~1\,$\mu$m, where more laser power is available, are more suitable for this purpose. For example, a tweezer at 1070\,nm with 40\,W power and a waist of $2.5\,\mu$m, generates a tweezer trap frequency of $\omega_\text{tw}\sim 2\pi \times 200$\,kHz~\cite{Espinoza:2021} and a further reduction in waist size can be achieved by increasing the NA of the tweezer delivery system. After performing the pre-alignment routine discussed above, these non-resonant tweezers can be aligned on the ions using one of the following options. 
A very precise method is to detect the Stark shift due to the tweezer on a clock transition~\cite{Haffner:2003a}. 
For instance in Ref.~\cite{Hirzler2020esf}, a high power 1070\,nm optical dipole trap was aligned on a trapped ion by detecting the tiny differential Stark shift on the hyperfine $S_{1/2}$ clock states in $^{171}$Yb$^+$. 
However, running the full optimization routine with this scheme is too time-consuming. One may instead use the impact of a far-detuned optical tweezer on the fluorescence of the ion during Doppler cooling as a means to align the tweezer. 
Since the tweezer shifts the transitions out of resonance, one can align and optimize the tweezers by minimizing the fluorescence. In the absence of a sufficiently large differential Stark shift between the states involved, one can instead align the tweezer by optimizing the shift in trap frequency of the ions.

Finally, we note that in the current setup, the tweezer is not actively stabilized.
Still, throughout the day, the tweezers' position relative to the ion varies by a maximum of $0.5(1)\,\mu$m, while over the span of several weeks, it fluctuates by $1.8(3)\,\mu$m. This effect can be further compensated by active beam pointing stabilization~\cite{Drechsler:2021}.

 









\section*{Acknowledgements}
We gratefully acknowledge discussions with Robert Spreeuw and support from Thomas Feldker and Henrik Hirzler in the construction stage of the setup. This work was supported by the Netherlands Organization for Scientific Research (Grant Nos. 680.91.120, VI.C.202.051 and 680.92.18.05, R.G., M.M. and R.X.S.). A.S.N is supported by the Dutch Research Council (NWO/OCW) as part of the Quantum Software Consortium programme (project number 024.003.037). A.S.N. and B.G. are supported by Quantum Delta NL (project number NGF.1582.22.030).





\appendix
\section{Modelling the ion fluorescence in the tweezer}
\label{appn:A}

In order to model the scattering behaviour of the ion 
we label the states in the 4-level model as follows:
$\ket{S}=S_{1/2}$, $\ket{P}=P_{1/2}$, $\ket{D}=D_{3/2}$ and $\ket{3/2}=[3/2]_{1/2}$.
Under the rotating wave approximation and in the interaction picture, the 4-level Hamiltonian is given by
\begin{equation}
\resizebox{0.9\columnwidth}{!}{$
\hat{\mathcal{H}}_{\rm tw}=\Delta_\text{D} \ketbra{S}{S} + \frac{1}{2} \Omega_{\text{D}}\hat{\sigma}_{x}^{S,P} + \Delta_\text{T} \ketbra{D}{D} + \frac{1}{2}\Omega_\text{T}(\mathbf{r})\hat{\sigma}_{x}^{D,3/2},
$}\label{eqn:H4_level}
\end{equation}
where $\Delta_\text{D}$ and $\Omega_{\text{D}}$ are the detuning and Rabi frequency on the Doppler transition, $\Delta_\text{tw}$ and $\Omega_{\text{tw}}(\mathbf{r})$ are the detuning and spatially-dependent Rabi frequency of the tweezer and $\hat{\sigma}_{x}^{i,j}=\ketbra{i}{j}+ \text{h.c.}$ is an operator driving between levels $\ket{i}$ and $\ket{j}$.

We account for the finite lifetime of the excited states by introducing a jump operator for each of the four relevant decay paths, which fall into two branches: (i) decay from $\ket{P}$ to $\ket{S}$ or $\ket{D}$ given by $\hat{c}_{f,P}=\sqrt{\Gamma_{f,P}} \ketbra{f}{P}$, and (ii) decay from $\ket{3/2}$ to $\ket{S}$ or $\ket{D}$ given by  $\hat{c}_{f,3/2}=\sqrt{\Gamma_{f,3/2}} \ketbra{f}{3/2}$ with $\ket{f}=\ket{S},\ket{D}$. The decay rates are given by $\Gamma_{f,P}= \Gamma_{\text{P}} b_{f,P}$, $\Gamma_{f,3/2}= \Gamma_{3/2} b_{f,3/2}$ with corresponding linewidths $\Gamma_{\text{P}}/2 \pi= 21$\,MHz, $\Gamma_{3/2}/2 \pi= 4.2$\,MHz, and branching ratios $b_{S,P}=0.995$, $b_{3/2,P}=0.005$, $b_{S,P}=0.018$, $b_{3/2,P}=0.982$~\cite{Olmschenk:2007}.

We use the Hamiltonian in Eq.~\eqref{eqn:H4_level} and the jump operators above in the Lindblad master equation, 
$
   \dot \rho = -i \left[\hat{\mathcal{H}}_{\rm tw} , \rho \right] + \sum_{i=1}^4\left(\hat c_{i} \rho \hat c_{i}^\dagger - \frac12\lbrace \hat c_{i}^\dagger \hat c_{i},\rho\rbrace \right)
$
where $\hat c_i$ denotes the $i$-th jump operator. We solve for the steady state in order to obtain an expression for the scattering rate on the Doppler transition and find, 
\begin{equation}
\resizebox{0.9\columnwidth}{!}{$
    \Gamma_{\rm sc}=\frac{\Gamma_{\text{S,}3/2} \Gamma_{\text{P}} \Omega_{\rm D}^2 \Omega_{\rm tw}^2}{\Gamma_{\text{S,}3/2} \Omega_{\rm tw}^2 \left( \Gamma_{\text{P}}^2 + 2 \Omega_{\rm D}^2 + 4 \Delta_{\rm D}^2 \right)+ \Gamma_{\text{S,P}} \Omega_{\rm D}^2 \left( \Gamma_{3/2}^2 + 2 \Omega_{\rm tw}^2 + 4 \Delta_{\rm tw}^2 \right)},
    $}
\label{eqn:4levelscatteringAPP}
\end{equation}
which we use to extract the tweezer waist by explicitly substituting a position dependent $\Omega_{\rm tw}(\mathbf{r})$.

The 4-level model is not sufficient to explain the coherent population trapping observed in the experiment. To this end, we use a 10-level description to account for the effects of Zeeman shifts and coherent population trapping.
In the 10-level description, the tweezer Hamiltonian is given by
\begin{equation}
\resizebox{\columnwidth}{!}{%
$\begin{aligned}
    \hat{\mathcal{H}}_{\rm tw, 10} &= (\Delta_\text{D} + \Delta_{Z,x}^{S_n}) \ketbra{S_n}{S_n} + \frac{1}{2} \left(\Omega_{\text{D}}^{n,m} \ketbra{S_n}{P_m} + h.c. \right) \\
    &+   \Delta_{Z,x}^{P_n} \ketbra{P_n}{P_n} \\
    &+ (\Delta_\text{tw}+ \Delta_{Z,x}^{D_n}) \ketbra{D_n}{D_n} + \frac{1}{2} \Omega_\text{tw}^{n,m}(\mathbf{r}) \left( \ketbra{{D_n}}{3/2_m} +h.c.\right)\\
    &+ \Delta_{Z,x}^{3/2_n} \ketbra{3/2_n}{3/2_n},
    \label{eqn:H10_level}
\end{aligned}$%
}
\end{equation}
where we have considered two $m_J$ sub-levels for the $\ket{S}$ states, two for the $\ket{P}$ states, four for the $\ket{D_{3/2}}$ states, and two for the $\ket{[3/2]_{1/2}}$ states. In our notation, the summation over indices $n$ and $m$ is implicit. 
The Rabi frequencies are now polarisation dependent and are defined as $\Omega_{\rm D}^{n,m}=\sum_q \boldsymbol{\epsilon}_{\rm d}\cdot \boldsymbol{\xi}_q\mathcal{A}^{\lambda}_{mJ_m,mJ_n} \Omega_{\rm D}$ and $\Omega_{\rm tw}^{n,m}=\sum_q \boldsymbol{\epsilon}_{\rm tw}\cdot \boldsymbol{\xi}_q\mathcal{A}^{\lambda}_{mJ_m,mJ_n} \Omega_{\rm tw}$. 
\\

The values of the $\mathcal{A} $ coefficients used for the $369$\,nm  Doppler transitions are
$(\mathcal{A}^{369}_{\pm 1/2, \mp 1/2},\mathcal{A}^{369}_{\pm 1/2,\pm 1/2})= (\mp \frac{1}{3},\frac{\sqrt{3}}{3})$. For the $935$\,nm tweezer transition we use 
 $(\mathcal{A}^{935}_{\pm 3/2, \pm 1/2},\mathcal{A}^{935}_{\pm 1/2, \pm 1/2},\mathcal{A}^{935}_{\pm 1/2, \mp 1/2})= (- \frac{1}{\sqrt{3}}, \frac{\sqrt{2}}{3},\frac{1}{3})$.
The polarisation basis vectors are given by $\boldsymbol{\xi}_{q=\pm 1, 0}= , \mp\frac{1}{\sqrt{2}} \left( \boldsymbol{u_x} \pm i \boldsymbol{u}_y\right),\boldsymbol{u}_z$. Finally, the polarisations of the Doppler laser and tweezer laser are given by $\boldsymbol{\epsilon}_{\rm D}= \boldsymbol{u}_z$ and $\boldsymbol{\epsilon}_{\rm tw}= \boldsymbol{u}_y$.
The linear Zeeman shifts are given by $\Delta_{Z,x}=g_J \mu_\text{B} m_J B$, where $g_J$ is the Landé g-factor, the magnetic field is along the $x-$direction \cite{NIST_ASD} and $\mu_\text{B}$ is the Bohr magneton.

In the 10-level description we include all dipole allowed decay paths which stay within the 10 level manifold. The corresponding jump operators are
\begin{equation}
    \begin{split}
        \hat{c}_{f_n,P_m}&=\sqrt{\Gamma_{f_n,P_m}} \ketbra{f_n}{P_m} \\
        \hat{c}_{f_n,3/2_m}&=\sqrt{\Gamma_{f_n,3/2_m}} \ketbra{f_n}{3/2_m}, \\
    \end{split}
    \label{eqn:jumps_10level}
\end{equation}
where $\ket{f_n}=\ket{S_n},\ket{D_n}$ is the final state in one of the Zeeman manifolds. The decay rates $\Gamma$ are modified from the 4-level case and are given by $\Gamma'=\Gamma (\mathcal{A}^{\lambda}_{m',m})^2$.

\section{Modelling mechanical effects in the tweezer}
\label{appn:B}
A simple model is used to simulate the effect of radiation forces on the ion.
The model includes three forces;
the force provided by the Paul trap, which is given by 
\begin{equation}
    F_\text{P}(z)=-m \omega_{\text{z}}^2 z,
\end{equation}
where $m$ is the mass of the ion, $ \omega_{\text{z}}$ is the axial trapping frequency and $z $ is the position of the ion.
The tweezer potential is assumed to be given by $U=\hbar \Omega_T^2(z)/\Delta_T$ multiplied by the population in the trapped $\ket{D_{3/2}}$ state. The tweezer force is then given by
\begin{equation}
    F_{\text{tw}}(z)=-\frac{d}{dz} \frac{ \Omega_\text{tw}(z)^2}{\Delta_\text{tw}} \rho_{D_{3/2}}(z),
\end{equation}
where $\rho_{D_{3/2}}$ is given by the steady state solution of the 4-level model in which we input 
\begin{equation}
    \Omega_\text{tw}(z)=\Omega_0 e^{\frac{-(z-z_{\text{tw}})^2}{w_0^2}},
    \label{eqn:gauss}
\end{equation} $z$ is the position of the ion and $z_\text{tw}$ is the position of the tweezer.
The scattering force provided by the Doppler beam is approximated as the momentum per photon multiplied by the scattering rate on the 369\,nm transition.
\begin{equation}
    F_{\text{D}}(z)=\hbar k \Gamma_{\text{P}} \rho_{P_{1/2}}(z),
\end{equation}
where $k=2\pi /\lambda$, $\lambda=369$\,nm and the population $\rho_{P_{1/2}}(z)$ is once again given by the steady state solution of the 4-level model.
\\
We now solve $F_{\text{P}}(z)+F_{\text{tw}}(z)+F_{\text{D}}(z)=0$ for the ion equilibrium position $z$. 
This equilibrium position can be plugged into the expression for the scattering rate given in equation \ref{eqn:4levelscatteringAPP} with tweezer profile given by equation \ref{eqn:gauss} after which a scattering rate curve like in figure \ref{fig:rabifit} can be generated by changing the tweezer position $z_{\text{tw}}$. 
We now generate scattering rate curves for different $\omega_z$ and find the value of $z_{\text{tw}}=z_0$ at which the scattering rate is maximized for each value of $\omega_z$. 
The resulting $\omega_z, z_0$ curve can directly be compared to the positions of the peak scattering rate in the fitted experimental data and is shown in figure \ref{fig:rabifit}. 
We observe a clear $\propto \omega_z^{-2}$ trend in the theory line.

\section{SLM pattern}
\label{appn:C}
\begin{figure}[h]
    \includegraphics[trim={0.7cm 0 0 0},width=\linewidth]{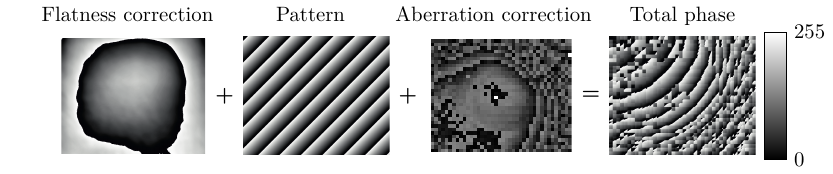}
    \caption{
    The total phase put on the SLM consists of the flatness correction given by the company, the diffraction grating pattern (or additional optical patterns), and the aberration correction for each zone from our algorithm.
    }
    \label{fig:SLM_pattern}
\end{figure}

The complete phase pattern rendered on the SLM consists of several components as shown in Fig.~\ref{fig:SLM_pattern}. First, a flatness correction pattern, supplied by the manufacturer, compensates for the SLM's lack of flatness. Second, a diffractive pattern steers the beam to converge at the ion's location this is also used to separate the SLM zero-th order diffraction. Finally, the aberration correction pattern, obtained as described in the main text, compensates for the optical aberrations of the system.

\section{Wavefront aberration identification}

Wavefront aberrations can be characterized using Zernike polynomials. These polynomials constitute a full orthogonal basis over a unit disk and each have a direct correspondence with a type of optical aberration. For this reason they are quite significant in fields such as optics and ophthalmology~\cite{Niu_2022,McAlinden:2011}. 

\begin{figure}[h]
    \centering
    \includegraphics[width= \columnwidth]{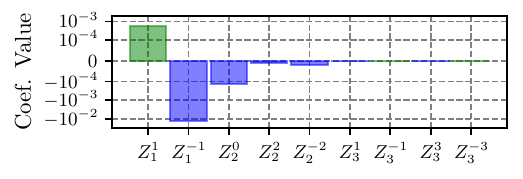}
    \caption{Obtained coefficients of the Zernike polynomial fit. The zeroth order polynomial is omitted.}
    \label{fig:Zernikedecomposition}
\end{figure}

We decompose our optimisation pattern into the Zernike polynomial basis by performing a polynomial fit and thus identify the main aberration types present in our system, figure \ref{fig:Zernikedecomposition}.

\begin{figure}[h]
    \centering
    \includegraphics[width= \columnwidth]{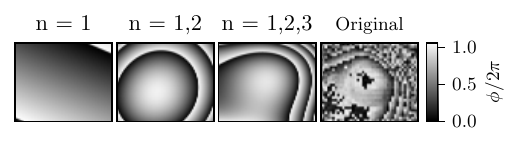}
    \caption{Wavefront aberration pattern decomposed into the first, second and third Zernike polynomial orders.}
    \label{fig:Zernikes}
\end{figure}

First-order polynomials account for vertical or horizontal tilts and primarily indicate positional changes. They constitute gratings added to the pattern, which indicate a small discrepancy between ion and tweezer position. The second-order polynomials correspond to defocusing and astigmatism aberrations. Third-order polynomials describe trefoil, which is deformation from three directions, and coma, which causes comet-like distortions from off-axis point sources. We observe that these initial polynomial orders capture the main features of the wavefront aberrations in our system and are thus crucial for their identification, figure \ref{fig:Zernikes}.

Further refinement of our correction pattern could be achievable by selecting an optimal combination of Zernike polynomial coefficients which can be independently optimised due to the orthogonality of their basis. Values derived from initial fittings provide useful starting points. Iteratively adjusting a few key polynomials has proven effective in significantly enhancing beam correction~\cite{Liang:2018,Werf:2017,Kontenis:2020}.


\bibliographystyle{apsrev4-2}
\bibliography{paper}

\end{document}